\documentclass[pre,aps,preprint,showpacs]{revtex4-1}
\usepackage[T1]{fontenc}
\usepackage{revsymb4-1,latexsym,amssymb,amsmath,comment}
\usepackage{graphicx,psfrag,subfigure,stmaryrd,xcolor}
\usepackage[ansinew]{inputenc}

\newcommand{\be}{\begin{equation}}
\newcommand{\ee}{\end{equation}}
\newcommand{\bel}[1]{\begin{equation}\label{#1}}
\newcommand{\bea}{\begin{eqnarray}}
\newcommand{\eea}{\end{eqnarray}}
\newcommand{\ba}{\begin{array}}
\newcommand{\ea}{\end{array}}
\newcommand{\bef}{\begin{figure}}
\newcommand{\ef}{\end{figure}}

\begin{document}

\author{Thomas Bose and Steffen Trimper}
\affiliation{Institute of Physics,
Martin-Luther-University, D-06099 Halle, Germany}
\email{thomas.bose@physik.uni-halle.de}
\email{steffen.trimper@physik.uni-halle.de}
\title{Nonlocal feedback in ferromagnetic resonance}
\date{\today }

\begin{abstract}
Ferromagnetic resonance in thin films is analyzed under the influence of spatiotemporal feedback effects.
The equation of motion for the magnetization dynamics is nonlocal in both space and time and includes
isotropic, anisotropic and dipolar energy contributions as well as the conserved Gilbert- and the non-conserved
Bloch-damping. We derive an analytical expression for the peak-to-peak linewidth. It consists of four
separate parts originated by Gilbert damping, Bloch-damping, a mixed Gilbert-Bloch component and a contribution
arising from retardation. In an intermediate frequency regime the results are comparable with the commonly
used Landau-Lifshitz-Gilbert theory combined with two-magnon processes. Retardation effects together with Gilbert damping lead
to a linewidth the frequency dependence of which becomes strongly nonlinear. The relevance and the applicability of our approach
to ferromagnetic resonance experiments is discussed.


\end{abstract}

\maketitle

\section{Introduction}

Ferromagnetic resonance enables the investigation of spin wave damping in thin or ultrathin ferromagnetic films.
The relevant information is contained in the linewidth of the resonance signal \cite{Heinrich:Kapitel3UtrathinMagStrucII:2005,Heinrich:UltrathinMagStrucIII:143:2005,MillsRezende:SpinDamping:SpinDynConfMagStruc:2003}.
Whereas the intrinsic damping included in the Gilbert or Landau-Lifshitz-Gilbert equation \cite{Landau:ZdS:8:p153:1935,Gilbert:ITOM:40:p3443:2004}, respectively, predicts
a linear frequency dependence of the linewidth ~\cite{Celinski:JMagMagMat166:6:1997}, the extrinsic
contributions associated with two-magnon scattering processes show a nonlinear behavior. Theoretically two-magnon scattering was analyzed for the case that the static external field lies in the film plane
\cite{Arias:PhysRevB60:7395:1999,Arias:JAP87:5455:2000}. The theory was quantitatively validated by experimental investigations with regard to the film
thickness \cite{Azevedo:PRB62:5331:2000}. Later the approach was extended to the case of arbitrary angles between the external field and the film surface
\cite{Landeros:PhysRevB77:214405:2008}. The angular dependence of the linewidth is often modeled by a sum of contributions including angular spreads and internal
field inhomogeneities \cite{Chappert:PRB34:3192:1986}. Among others, two-magnon mechanisms were used to explain the experimental observations
\cite{Hurben:JApllPhys81:7458:1997,McMichael:JApplPhys83:7037:1998,Woltersdorf:PhysRevB69:184417:2004,Krivosik:ApplPhysLett95:052509:2009,
Lindner:PRB80:224421:2009,Dubowik:PRB84:184438:2011} whereas the influence of the size of the inhomogeneity was studied in
\cite{McMichael:PRL90:227601:2003}. As discussed in \cite{MillsRezende:SpinDamping:SpinDynConfMagStruc:2003,Woltersdorf:PhysRevB69:184417:2004} the
two-magnon contribution to the linewidth disappears for tipping angles between magnetization and film plane exceeding a critical one $\Phi_{\rm{M}}^{crit}=\pi /4$.
Recently, deviations from this condition were observed comparing experimental data and numerical
simulations \cite{Dubowik:PRB84:184438:2011}.
Spin pumping can also contribute to the linewidth  as studied theoretically in \cite{Costa:PRB73:054426:2006}. However, a superposition of both the Gilbert damping
and the two-magnon contribution turned out to be in agreement very well with experimental data illustrating the dependence of the linewidth on the frequency \cite{Lindner:PhysRevB68:060102:2003,Lenz:PhysRevB73:144424:2006,Zakeri:PRB76:2007:104416,Zakeri:PhysRevB80:059901:2009,Lindner:PRB80:224421:2009}. Based on these findings it was put into question whether the Landau-Lifshitz-Gilbert
equation is an appropriate description for ferromagnetic thin films. The pure Gilbert damping is not
able to explain the nonlinear frequency dependence of the linewidth when two-magnon scattering processes are
operative \cite{MillsRezende:SpinDamping:SpinDynConfMagStruc:2003,HoMaAMM:Baberschke:2007}. Assuming that damping mechanisms can also lead to a non-conserved spin length a way out might be the inclusion of the Bloch
equations \cite{Bloch:PhysRev70:460:1946,Bloembergen:PhysRev78:572:1950} or the the Landau-Lifshitz-Bloch equation
\cite{Garanin:TheoMatPhys82:169:1990,Garanin:PRB:55:p3050:1997} into the concept of ferromagnetic resonance.

Another aspect is the recent observation \cite{Barsukov:PRB84:140410:2011} that a periodic scattering potential
can alter the frequency dependence of the linewidth. The experimental results are not in agreement with those
based upon a combination of Gilbert damping and two-magnon scattering. It was found that the linewidth
as function of the frequency exhibits a non monotonous behavior. The authors \cite{Barsukov:PRB84:140410:2011}
suggest to reconsider the approach with regard to spin relaxations. 
Moreover, it would be an advantage to derive an expression for the linewidth as a measure for
spin damping solely from the equation of motion for the magnetization.

Taking all those arguments into account it is the aim of this paper to propose a generalized equation of
motion for the magnetization dynamics including both Gilbert damping and Bloch terms. The dynamical model
allows immediately to get the magnetic susceptibility as well as the ferromagnetic resonance linewidth
which are appropriate for the analysis of experimental observations. A further generalization is the
implementation of nonlocal effects in both space and time.
This is achieved by introducing a retardation kernel which takes into account temporal retardation within
a characteristic time $\tau$ and a spatial one with a characteristic scale $\xi$. The last one simulates
an additional  mutual interaction of the magnetic moments in different areas of the film within the retardation
length $\xi$. Recently such nonlocal effects were discussed in a complete different context
\cite{PhysRevLett.108.057204}. Notice that retardation effects were already investigated for simpler models
by means of the Landau-Lifshitz-Gilbert equation. Here the existence of spin wave solutions were in the focus
of the consideration \cite{bosetrimper:retardatin:PRB:2011}.
The expressions obtained for the frequency/damping parameters were converted into linewidths according to the
Gilbert contribution which is a linear function of the frequency
\cite{bosetrimper:retardatin:PRB:2011,BoseTrimper:pssb:2011}. In the present approach we follow another line.
The propagating part of the varying magnetization is supplemented by the two damping terms due to Gilbert and
Bloch, compare Eq.~\eqref{eom1}. Based on this equation we derive analytical expressions
for the magnetic susceptibility, the resonance condition and the ferromagnetic resonance linewidth.
Due to the superposition of damping and retardation effects the linewidth exhibits a nonlinear behavior
as function of the frequency.
The model is also extended by considering the general case of arbitrary angles between the static external field
and the film surface. Moreover the model includes several energy contributions as Zeeman and exchange energy as
well as anisotropy and dipolar interaction. The consequences for ferromagnetic resonance experiments are
discussed.

\section{Derivation of the equation of motion}

In order to define the geometry considered in the following we adopt the idea presented in \cite{Landeros:PhysRevB77:214405:2008}, i.e. we employ two coordinate systems,
the $xyz$-system referring to the film surface and the $\rm{XYZ}$-system which is canted by an angle $\Theta_{\rm{M}}$ with respect to the film plane. The situation
for a film of thickness $d$ is sketched in Fig.~\ref{filmkos}.
\bef
\includegraphics[width=7.5cm]{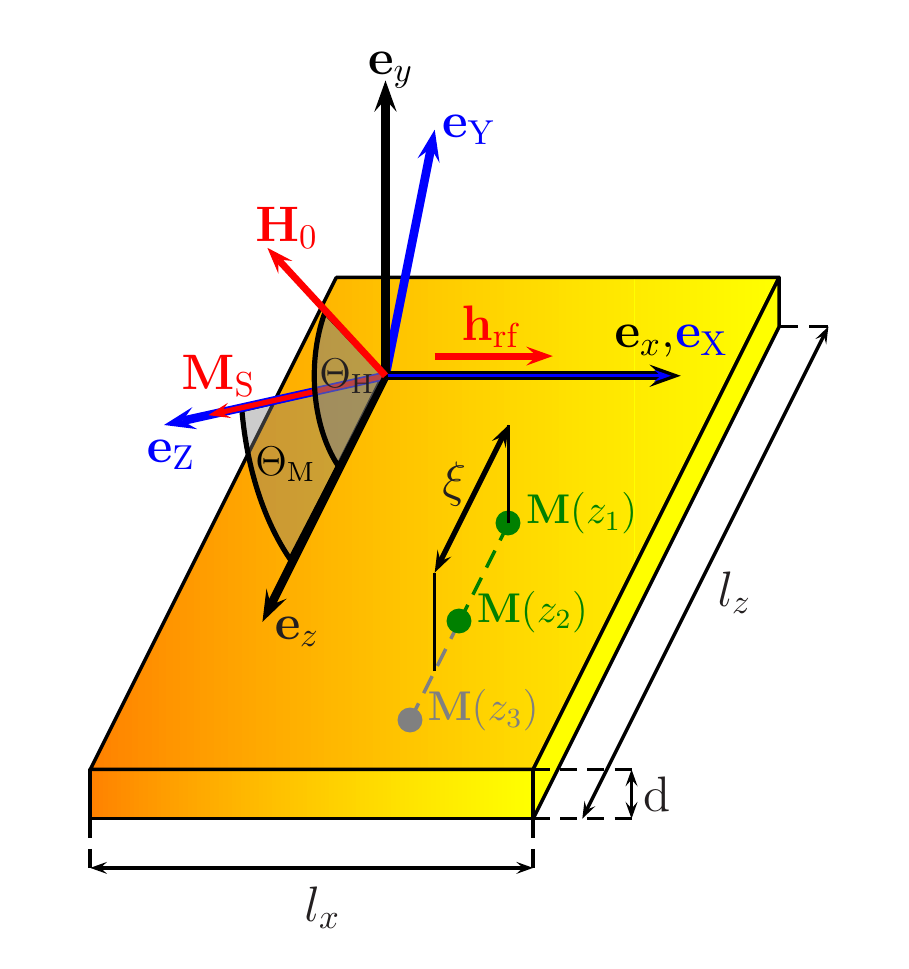}
\caption{(Color online) The geometry referring to the film and the magnetization. Further description in the text.}
\label{filmkos}
\ef
The angle $\Theta_{\rm{M}}$ describing the direction of the saturation magnetization, aligned with the $\rm{Z}$-axis, originates from the static external field
$\mathbf{H}_0$ which impinges upon the film surface under an angle $\Theta_{\rm{H}}$. Therefore, it is more convenient to use the $\rm{XYZ}$-system for the
magnetization dynamics. As excitation source we consider the radio-frequency (rf) magnetic field $h_{\rm{rf}}$ pointing into the $x=\rm{X}$-direction. It should fulfill the condition
$h_{\rm{rf}}\ll H_0$. To get the evolution equation of the magnetization $\mathbf{M}(\mathbf{r},t)$, $\mathbf{r}=(x,y,z)$ we have to define the energy of the system.
This issue is well described in Ref.~\cite{Landeros:PhysRevB77:214405:2008}, so we just quote the most important results given there and refer to the cited literature for details.
Since we consider the thin film limit one can perform the average along the direction perpendicular to the film, i.e.
\begin{align}
\begin{aligned}
\mathbf{M}(\mathbf{r}_\parallel ,t)=\frac{1}{d}\,\int_{-d/2}^{d/2}\mathrm{d}y\, \mathbf{M}(\mathbf{r},t) \,,
\end{aligned}
\label{magaverage}
\end{align}
where $\mathbf{r}_\parallel =(x,0,z)$ lies in the film plane. In other words the spatial variation of the magnetization across the film thickness $d$ is neglected.
The components of the magnetization point into the directions of the $\rm{XYZ}$-system and can be written as \cite{Gurevich:book:1996:MagOsWaves}
\begin{align}
\begin{aligned}
\mathbf{M}(\mathbf{r}_\parallel ,t)=M_{\rm{X}}(\mathbf{r}_\parallel)\,\mathbf{e}_{\rm{X}} + M_{\rm{Y}}(\mathbf{r}_\parallel)\,\mathbf{e}_{\rm{Y}} + \left(M_{\rm{S}}-\frac{M^2_{\rm{X}}
(\mathbf{r}_\parallel)+M^2_{\rm{Y}}(\mathbf{r}_\parallel)}{2\,M_{\rm{S}}}\right)\,\mathbf{e}_{\rm{Z}} \,.
\end{aligned}
\label{magexpand}
\end{align}
Typically the transverse components $M_{\rm{X},\rm{Y}}$ are assumed to be much smaller than the saturation magnetization $M_{\rm{S}}$. Remark that terms quadratic in
$M_{\rm{X},\rm{Y}}$ in the energy will
lead to linear terms in the equation of motion. The total energy of the system can now be expressed in terms of the averaged magnetization from Eq.~\eqref{magaverage} and reads
\begin{align}
\begin{aligned}
\mathcal{H}=\mathcal{H}_{\rm{z}} + \mathcal{H}_{\rm{ex}} + \mathcal{H}_{\rm{a}} + \mathcal{H}_{\rm{d}} \,.
\end{aligned}
\label{totalenergy}
\end{align}
The different contributions are the Zeeman energy
\begin{align}
\begin{aligned}
\mathcal{H}_{\rm{z}} = &-\int \mathrm{d}^3r \,H_0\sin\left(\Theta_{\rm{H}}-\Theta_{\rm{M}}\right)M_{\rm{Y}}(\mathbf{r}_\parallel)\\
&-\int \mathrm{d}^3r \,H_0\cos\left(\Theta_{\rm{H}}-\Theta_{\rm{M}}\right) \left(M_{\rm{S}}-\frac{M_{\rm{X}}(\mathbf{r}_\parallel)^2+M_{\rm{Y}}(\mathbf{r}_\parallel)^2}{2\,M_{\rm{S}}}\right) \,,
\end{aligned}
\label{energyzeeman}
\end{align}
the exchange energy
\begin{align}
\begin{aligned}
\mathcal{H}_{\rm{ex}} = \frac{D}{2 M_{\rm{S}}}\, \int \mathrm{d}^3r \,\left[\boldsymbol\nabla M_{\rm{X}}(\mathbf{r}_\parallel)\right]^2 + \left[\boldsymbol\nabla M_{\rm{Y}}
(\mathbf{r}_\parallel)\right]^2 \,, \\
\end{aligned}
\label{energyexchange}
\end{align}
the surface anisotropy energy
\begin{align}
\begin{aligned}
\mathcal{H}_{\rm{a}}= &\frac{H_{\rm{S}} M_{\rm{S}} V}{2}\,\sin^2\left(\Theta_{\rm{M}}\right) + \frac{H_{\rm{S}}}{2}\,\sin(2\Theta_{\rm{M}})\int \mathrm{d}^3r \,M_{\rm{Y}}(\mathbf{r}_\parallel) \\
&+\frac{H_{\rm{S}}}{2\,M_{\rm{S}}}\,\cos(2\Theta_{\rm{M}})\int \mathrm{d}^3r\,M_{\rm{Y}}(\mathbf{r}_\parallel)^2 - \sin^2(\Theta_{\rm{M}})\int
\mathrm{d}^3r\,M_{\rm{X}}(\mathbf{r}_\parallel)^2 \,, \\
\end{aligned}
\label{energyanisotropy}
\end{align}
and the dipolar energy
\begin{align}
\begin{aligned}
\mathcal{H}_{\rm{d}} = &2\pi M_{\rm{S}}^2 V\,\sin^2(\Theta_{\rm{M}}) + \pi \int \mathrm{d}^3r \,\bigg\{ 2M_{\rm{S}}\sin(2\Theta_{\rm{M}})\,M_{\rm{Y}}(\mathbf{r}_\parallel) \\
 &{}+\left( \frac{dk_z^2}{k_\parallel}\,\sin^2(\Theta_{\rm{M}})-(dk_\parallel -2)\,\cos^2(\Theta_{\rm{M}})-2\,\sin^2(\Theta_{\rm{M}}) \right)\,M_{\rm{Y}}(\mathbf{r}_\parallel)^2 \\
 &{}+\left(\frac{dk_x^2}{k_\parallel}-2\,\sin^2(\Theta_{\rm{M}})\right)\,M_{\rm{X}}(\mathbf{r}_\parallel)^2-\frac{2dk_xk_z}{k_\parallel}\,\sin(\Theta_{\rm{M}})\,
 M_{\rm{X}}(\mathbf{r}_\parallel) M_{\rm{Y}}(\mathbf{r}_\parallel)  \bigg\} \,.
\end{aligned}
\label{energydipol}
\end{align}
In these expressions $V=l_xl_zd$ is the volume of the film, $D$ designates the exchange stiffness and $H_{\rm{S}}\propto d^{-1}$ represents the uniaxial out-of-plane anisotropy field.
If $H_{\rm{S}}<0$ the easy axis is perpendicular to the film surface. The in-plane anisotropy contribution to the energy is neglected but it should be appropriate for
polycrystalline samples \cite{Lindner:PRB80:224421:2009}.
Moreover $k_\parallel=|\mathbf{k}_\parallel|$ is introduced where $\mathbf{k}_\parallel=k_x\,\mathbf{e}_x+k_z\,\mathbf{e}_z$ is the wave vector of the spin waves parallel to the film surface. Eqs.~\eqref{totalenergy}-\eqref{energydipol} are valid in the thin film limit $k_\parallel d\ll 1$.
In order to derive $\mathcal{H}_{\rm{d}}$ in Eq.~\eqref{energydipol} one defines a scalar magnetic potential and has to solve the corresponding boundary value problem
inside and outside of the film \cite{DamonEshbach:JPhysChemSol19:308:1961}. As result \cite{Landeros:PhysRevB77:214405:2008} one gets the expressions in Eq.~\eqref{energydipol}.

In general if the static magnetic field is applied under an arbitrary angle $\Theta_{\rm{H}}$ the magnetization does not align in parallel, i.e. $\Theta_{\rm{M}}\neq \Theta_{\rm{H}}$.
The angle $\Theta_{\rm{M}}$ can be derived from the equilibrium energy $\mathcal{H}_{\rm{eq}}=\mathcal{H}(M_{\rm{X}}=0,M_{\rm{Y}}=0)$.
Defining the equilibrium free energy density as $f_{\rm{eq}}(\Theta_{\rm{M}})=\mathcal{H}_{\rm{eq}}/V$ according to Eqs.~\eqref{totalenergy}-\eqref{energydipol}
one finds the well-known condition
\begin{align}
\begin{aligned}
\sin(\Theta_{\rm{H}}-\Theta_{\rm{M}})=\frac{4\pi M_{\rm{S}}+H_{\rm{S}}}{2\,H_0}\,\sin(2\,\Theta_{\rm{M}})
\end{aligned}
\label{zerotorque}
\end{align}
by minimizing $f_{\rm{eq}}$ with respect to $\Theta_{\rm{M}}$. We further note that all terms linear in $M_{\rm{Y}}$ in Eqs.~\eqref{totalenergy}-\eqref{energydipol} cancel mutually
by applying Eq.~\eqref{zerotorque} as already pointed out in Ref.~\cite{Landeros:PhysRevB77:214405:2008}.

The energy contributions in Eqs.~\eqref{totalenergy} and the geometric aspects determine the dynamical equation for the magnetization.
The following generalized form is proposed
\begin{align}
\begin{aligned}
\frac{\partial}{\partial t}\mathbf{M}(\mathbf{r}_\parallel ,t) = \iint \mathrm{d}\mathbf{r}_\parallel'\mathrm{d}t' \;&\Gamma(\mathbf{r}_\parallel-\mathbf{r}_\parallel';t-t')\,\Bigg\{ \gamma\,\left[\mathbf{H}_{\textrm{eff}}(\mathbf{r}_\parallel',t')\times \mathbf{M}(\mathbf{r}_\parallel',t')\right] \\
&{}+\alpha\,\left[\mathbf{M}(\mathbf{r}_\parallel',t')\times \frac{\partial}{\partial t'}\mathbf{M}(\mathbf{r}_\parallel',t')\right]
-\frac{1}{T_2}\,\mathbf{M}_\perp (\mathbf{r}_\parallel',t')\Bigg\} \,,
\end{aligned}
\label{eom1}
\end{align}
where $\gamma=g\mu_B/\hbar$ is the absolute value of the gyromagnetic ratio, $T_2$ is the transverse relaxation time of the components
$\mathbf{M}_\perp =M_{\rm{X}}\,\mathbf{e}_{\rm{X}} + M_{\rm{Y}}\,\mathbf{e}_{\rm{Y}}$ and $\alpha$ denotes the dimensionless Gilbert damping
parameter. The latter is often transformed into $G=\alpha \gamma M_{\rm{S}}$ representing the corresponding damping constant in unit $\mathrm{s}^{-1}$.
The effective magnetic
field $\mathbf{H}_{\textrm{eff}}$ is related to the energy in Eqs.~\eqref{totalenergy}-\eqref{energydipol} by means of variational principles
\cite{MacdonaldProcPhysSoc64:968:1951}, i.e. $\mathbf{H}_{\textrm{eff}}=-\delta\mathcal{H}/\delta\mathbf{M}+\mathbf{h}_{\textrm{rf}}$.
Here the external rf-field $\mathbf{h}_{\rm{rf}}(t)$ is added which drives the system out of equilibrium.

Regarding the equation of motion presented in Eq.~\eqref{eom1} we note that a similar type was applied in \cite{Hurben:JApllPhys81:7458:1997} 
for the evaluation of ferromagnetic resonance experiments.
In this paper the authors made use of a superposition of the Landau-Lifshitz equation and Bloch-like relaxation.
Here we have chosen the part which conserves
the spin length in the Gilbert form and added the non-conserving Bloch term in the same manner.
That the combination of these two distinct damping mechanisms is suitable for the investigation of ultrathin magnetic films
was also suggested in \cite{HoMaAMM:Baberschke:2007}.
Since the projection of the magnetization onto the $Z$-axis is not affected by $T_2$ this relaxation time characterizes the transfer
of energy into the transverse components of the magnetization. 
This damping type is supposed to account for spin-spin relaxation processes such as magnon-magnon scattering \cite{Gurevich:book:1996:MagOsWaves,Suhl:IEEE:1998:34:1834}.
In our ansatz we introduce another possible source of damping by means of the 
feedback kernel $\Gamma(\mathbf{r}_\parallel-\mathbf{r}_\parallel';t-t')$.
The introduction of this quantity
reflects the assumption that the magnetization $\mathbf{M}(\mathbf{r}_\parallel,t_2)$ is not independent of its previous value $\mathbf{M}(\mathbf{r}_\parallel,t_1)$
provided $t_2-t_1<\tau$. Here $\tau$ is a time scale where the temporal memory is relevant. In the same manner the spatial feedback controls the magnetization dynamics
significantly on a characteristic length scale $\xi$, called retardation length. Physically, it seems to be reasonable that the retardation length differs
noticeably from zero only in $z$-direction which is shown in Fig.~\ref{filmkos}. As illustrated in the figure $\mathbf{M}(x,z_1,t)$ is affected by $\mathbf{M}(x,z_2,t)$
while $\mathbf{M}(x,z_3,t)$ is thought to have negligible influence on $\mathbf{M}(x,z_1,t)$ since $|z_3-z_1|>\xi$.
Therefore we choose the following combination of a local and a nonlocal part as feedback kernel
\begin{align}
\begin{aligned}
\Gamma(\mathbf{r}_\parallel-\mathbf{r}_\parallel';t-t') = &\Gamma_0\,\delta(\mathbf{r}_\parallel-\mathbf{r}_\parallel')\,\delta(t-t') \\
&{}+\frac{\Gamma_0}{4\,\xi\,\tau}\,\delta(x-x')\,\exp\left[\frac{-|z-z'|}{\xi}\right]\,\exp\left[\frac{-(t-t')}{\tau}\right] \,,\, t>t' \,.
\end{aligned}
\label{memorykernel}
\end{align}
The intensity of the spatiotemporal feedback is controlled by the dimensionless retardation strength $\Gamma_0$. The explicit form in Eq.~\eqref{memorykernel} is
chosen in such a manner that the Fourier-transform $\Gamma(\mathbf{k}_\parallel,\omega)\to \Gamma_0$ for $\xi\to 0$ and $\tau\to 0$, and in case $\Gamma_0=1$ the ordinary equation of motion for the magnetization is recovered.
Further, $\int{\mathrm{d}\mathbf{r}_\parallel\mathrm{d}t\,\Gamma(\mathbf{r}_\parallel,t)}=\Gamma_0<\infty$, i.e. the integral remains finite.

\section{Susceptibility and FMR-linewidth}

If the rf-driving field, likewise averaged over the film thickness, is applied in ${\rm{X}}$-direction,
i.e. $\mathbf{h}_{\textrm{rf}}(\mathbf{r}_\parallel,t)=h_{\rm{X}}(\mathbf{r}_\parallel,t)\,\mathbf{e}_{\rm{X}}$, the Fourier transform of Eq.~\eqref{eom1} is written as
\begin{align}
\begin{aligned}
\left[\frac{\textrm{i}\omega}{\gamma \,\Gamma(\mathbf{k}_\parallel,\omega)}+\frac{1}{\gamma \,T_2}+H_{21}(\mathbf{k}_\parallel)\right]\,M_{\rm{X}}(\mathbf{k}_\parallel,\omega) =& -\left[H_1(\mathbf{k}_\parallel)+\frac{\textrm{i}\alpha \omega}{\gamma}\right]\,M_{\rm{Y}}(\mathbf{k}_\parallel,\omega) \,,\\
\left[\frac{\textrm{i}\omega}{\gamma \,\Gamma(\mathbf{k}_\parallel,\omega)}+\frac{1}{\gamma \,T_2}+H_{12}(\mathbf{k}_\parallel)\right]\,M_{\rm{Y}}(\mathbf{k}_\parallel,\omega) =& \left[H_2(\mathbf{k}_\parallel)+\frac{\textrm{i}\alpha \omega}{\gamma}\right]\,M_{\rm{X}}(\mathbf{k}_\parallel,\omega)-M_{\rm{S}}\,h_{\rm{X}}(\mathbf{k}_\parallel,\omega) \,.
\end{aligned}
\label{fteom}
\end{align}
The effective magnetic fields are expressed by
\begin{align}
\begin{aligned}
H_1(\mathbf{k}_\parallel) =& H_0\,\cos(\Theta_{\rm{H}}-\Theta_{\rm{M}})+(4\pi M_{\rm{S}}+H_{\rm{S}})\,\cos(2\,\Theta_{\rm{M}}) \\
  & +2\pi dk_\parallel M_{\rm{S}}\,\left(\frac{k_z^2}{k_\parallel^2}\,\sin^2(\Theta_{\rm{M}})-\cos^2(\Theta_{\rm{M}})\right)+D\,k_\parallel^2 \\
H_2(\mathbf{k}_\parallel) =& H_0\,\cos(\Theta_{\rm{H}}-\Theta_{\rm{M}})-(4\pi M_{\rm{S}}+H_{\rm{S}})\,\sin^2(\Theta_{\rm{M}}) \\
  & +2\pi dM_{\rm{S}}\,\frac{k_x^2}{k_\parallel}+D\,k_\parallel^2   \,,
\end{aligned}
\label{H1H2}
\end{align}
and
\begin{align}
\begin{aligned}
H_{12}(\mathbf{k}_\parallel) = 2\pi d M_{\rm{S}}\,\frac{k_xk_z}{k_\parallel}\,\sin(\Theta_{\rm{M}}) = -H_{21}(\mathbf{k}_\parallel) \,.
\end{aligned}
\label{Hmix}
\end{align}
The Fourier transform of the kernel yields
\begin{align}
\begin{aligned}
\Gamma(\mathbf{k}_\parallel,\omega) =&\frac{\Gamma_0\,(1+\mathrm{i}\omega \tau)+\Gamma_1}{2\,(1+\mathrm{i}\omega \tau)}\; \stackrel{(\omega^2\tau^2\ll 1)}{\simeq}\; \frac{\Gamma_0+\Gamma_1}{2}-\frac{\mathrm{i}}{2}\,\Gamma_1\omega \tau \,, \\ \Gamma_1=&\frac{\Gamma_0}{1+\beta^2} \quad ,\quad \beta=\xi\,k_z \,,
\end{aligned}
\label{ftkernel}
\end{align}
where the factor $1/2$ arises from the condition $t>t'$ when performing the Fourier transformation from time into frequency domain. In Eq.~\eqref{ftkernel} we discarded
terms $\omega^2\tau^2\ll 1$. This condition is fulfilled in experimental realizations. So, it will be turned out later the retardation time $\tau\sim 10\,\mathrm{fs}$.
Because the ferromagnetic resonance frequencies are of the order $10 \dots100\,\mathrm{GHz}$ one finds $\omega^2\tau^2\sim 10^{-8}...10^{-6}$.
The retardation parameter $\beta=\xi k_z$, introduced in Eq.~\eqref{ftkernel}, will be of importance in analyzing the linewidth of the resonance signal.
With regard to the denominator in $\Gamma_1$, compare Eq.~\eqref{ftkernel}, the parameter
$\beta$ may evolve ponderable influence on the spin wave damping if this quantity cannot be neglected compared to $1$.
As known from two-magnon scattering the spin wave modes can be degenerated with the uniform resonance mode possessing wave vectors $k_\parallel \sim 10^5\,\mathrm{cm}^{-1}$.
The retardation length $\xi$ may be estimated by the size of inhomogeneities or the distance of defects on the film surface, respectively. Both length scales
can be of the order $\sim 10...1000\,\mathrm{nm}$, see Refs.~\cite{McMichael:PRL90:227601:2003,Barsukov:PRB84:140410:2011}. Consequently the retardation
parameter $\beta$ could reach or maybe even exceed the order of $1$.

Let us stress that in case $\beta=0$, $\tau=0$, $\Gamma_0=1$ and neglecting the Gilbert damping, i.e. $\alpha=0$, the spin wave dispersion relation
is simply $\gamma\sqrt{H_1(\mathbf{k}_\parallel)H_2(\mathbf{k}_\parallel)-H_{12}^2(\mathbf{k}_\parallel)}$.
This expression coincides with those ones given in Refs.~\cite{Arias:PhysRevB60:7395:1999} and \cite{Landeros:PhysRevB77:214405:2008}.

Proceeding the analysis of Eq.~\eqref{fteom} by defining the magnetic susceptibility $\chi$ as
\begin{align}
\begin{aligned}
M_\alpha(k_\parallel,\omega)=\sum_\beta\chi_{\alpha\beta}(k_\parallel,\omega)\,h_\beta(k_\parallel,\omega) \,,\qquad \{\alpha,\beta\}=\{{\rm{X}},{\rm{Y}}\} \,,
\end{aligned}
\label{defsusceptibility}
\end{align}
where $h_\beta$ plays the role of a small perturbation and the susceptibility $\chi_{\alpha\beta}$ exhibits the response of the system.
Eq.~\eqref{defsusceptibility} reflects that there appears no dependence on the direction of $\mathbf{k}_\parallel$.

Since the rf-driving field is applied along the $\mathbf{e}_{\rm{X}}$-direction it is sufficient to focus the following discussion
to the element $\chi_{\rm{XX}}^{\phantom{''}}$ of the susceptibility tensor. From Eq.~\eqref{fteom} we conclude
\begin{align}
\begin{aligned}
\chi_{\mathrm{XX}}^{\phantom{''}}(k_\parallel,\omega)=\frac{M_{\rm{S}}\left[H_1(k_\parallel,\omega)+\frac{\mathrm{i}\alpha\omega}{\gamma}\right]}
{\left[H_1(k_\parallel,\omega)+\frac{\mathrm{i}\alpha\omega}{\gamma}\right]\left[H_2(k_\parallel,\omega)+
\frac{\mathrm{i}\alpha\omega}{\gamma}\right]+\left[\frac{\mathrm{i}\omega}{\gamma\,\Gamma(k_\parallel,\omega)}+\frac{1}{\gamma T_2}\right]^2} \qquad .
\end{aligned}
\label{susceptibilityxx1}
\end{align}
Because at ferromagnetic resonance a uniform mode is excited let us set $k_\parallel=0$ in Eqs.~\eqref{H1H2}-\eqref{Hmix}. Considering the
resonance condition we can assume $\beta=\xi k_z=0$. For reasons mentioned above we have to take $\beta=\xi k_z\neq0$ when the linewidth as
a measure for spin damping is investigated. Physically we suppose that spin waves with non zero waves vectors are not excited at the moment of the
ferromagnetic resonance. However such excitations will evolve during the relaxation process. In finding the resonance condition from
Eq.~\eqref{susceptibilityxx1} it seems to be a reasonable approximation to disregard terms including the retardation time $\tau$. Such terms
give rise to higher order corrections. In the same manner all the contributions originated from the damping, characterized by $\alpha$ and $T_2$,
are negligible. Let us justify those approximation by quantitative estimations.
The fields $H_1$, $H_2$ and $\omega/\gamma$ are supposed to range in a comparable order of magnitude. On the other hand one finds
$\alpha \sim10^{-3}...10^{-2}$, $\omega T_2\sim10^{-2}$ and $\omega\tau \sim10^{-4}$. Under these approximations the resonance condition
reads
\begin{align}
\begin{aligned}
\left(\frac{\omega_\mathrm{r}}{\gamma}\right)^2=\Gamma_0^2\,H_1(k_\parallel=0)H_2(k_\parallel=0) \,.
\end{aligned}
\label{rescond}
\end{align}
This result is well known for the case without retardation with $\Gamma_0=1$. Although the retardation time $\tau$ and the retardation length $\xi$ are not incorporated in the resonance condition, the strength of the
feedback may be important as visible in Eq.~\eqref{rescond}. Now the consequences for the
experimental realization will be discussed. To address this issue the resonance condition
Eq.~\eqref{rescond} is rewritten in terms of the resonance field
$H_\mathrm{r}=H_0(\omega=\omega_\mathrm{r})$ leading to
\begin{align}
\begin{aligned}
H_\mathrm{r}=\frac{1}{2\,\cos(\Theta_{\rm{H}}-\Theta_{\rm{M}})}\left\{\sqrt{(4\pi M_{\rm{S}}+H_{\rm{S}})^2\cos^4(\Theta_{\rm{M}})+ \left(\frac{1}{\Gamma_0}\frac{2\,\omega_\mathrm{r}}{\gamma}\right)^2} \right.\\
\left. -(4\pi M_{\rm{S}}+H_{\rm{S}})(1-3\,\sin^2(\Theta_{\rm{M}})) \vphantom{\sqrt{\left(\frac{2\omega}{\gamma}\right)^2}}\right\} \,.
\end{aligned}
\label{resfield}
\end{align}
The result is arranged in the in the same manner as done in \cite{Lindner:PRB80:224421:2009}.
The difference is the occurrence of the parameter $\Gamma_0$ in the denominator.
In \cite{Lindner:PRB80:224421:2009} the gyromagnetic ratio $\gamma$ and the sum $(4\pi M_{\rm{S}}+H_{\rm{S}})$
were obtained from $\Theta_{\rm{H}}\,$-dependent measurements and a fit of the data according to
Eq.~\eqref{resfield} with $\Gamma_0=1$ under the inclusion of Eq.~\eqref{zerotorque}.
If the saturation magnetization can be obtained from other experiments \cite{Lindner:PRB80:224421:2009} the uniaxial anisotropy field $H_{\rm{S}}$ results.
Thus, assuming $\Gamma_0\neq 1$ the angular dependence $\Theta_{\rm{M}}(\Theta_{\rm{H}})$ and the fitting parameters as well would change. In Fig.~\ref{figphimphih} we illustrate the angle $\Theta_{\rm{M}}(\Theta_{\rm{H}})$
\bef
\includegraphics[width=7.5cm]{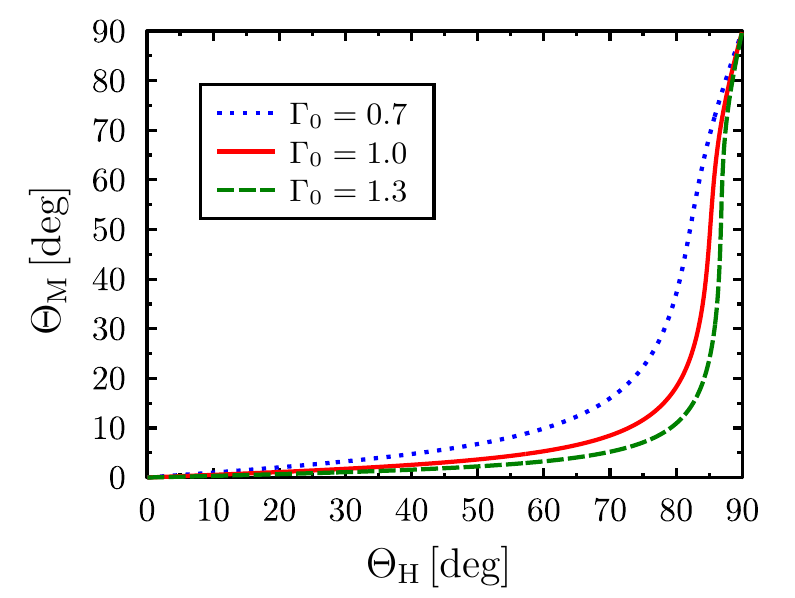}
\caption{(Color online) Dependence of the magnetization angle $\Theta_{\rm{M}}$ on the angle $\Theta_{\rm{H}}$
under which the static external field is applied for $\omega_\mathrm{r}/(2\pi)=10\,\mathrm{GHz}$. The parameters
are taken from ~\cite{Lindner:PRB80:224421:2009}:~$4\pi M_{\rm{S}}=16980\,\mathrm{G}$, $H_{\rm{S}}=-3400\,\mathrm{G},
\gamma=0.019\,\mathrm{GHz}/\mathrm{G}$.}
\label{figphimphih}
\ef
for different values of $\Gamma_0$ and a fixed resonance frequency. If $\Gamma_0 < 1$ the
curve is shifted to larger $\Theta_{\rm{M}}$ and for $\Gamma_0>1$ to smaller magnetization angles.
To produce Fig.~\ref{figphimphih} we utilized quantitative results presented in
\cite{Lindner:PRB80:224421:2009}. They found for Co films grown on GaAs the parameters
$4\pi M_{\rm{S}}=16980\,\mathrm{G}$, $H_{\rm{S}}=-3400\,\mathrm{G}$ and $\gamma=0.019\,\mathrm{GHz}/\mathrm{G}$.
As next example we consider the influence of $H_{\rm{S}}$ and denote $H_{\rm{S}}^{(0)}=-3400\,\mathrm{G}$ the anisotropy field for $\Gamma_0=1$ and $H_{\rm{S}}^{(R)}$ the anisotropy field for $\Gamma_0\neq1$. The absolute value of
their ratio $|H_{\rm{S}}^{(R)}/H_{\rm{S}}^{(0)}|$, derived from $H_\mathrm{r}(H_{\rm{S}}^{(0)},\Gamma_0=1)=H_\mathrm{r}(H_{\rm{S}}^{(R)},\Gamma_0\neq1)$, is depicted in Fig.~\ref{figanisotrpy} for various frequencies.
\bef
\includegraphics[width=7.5cm]{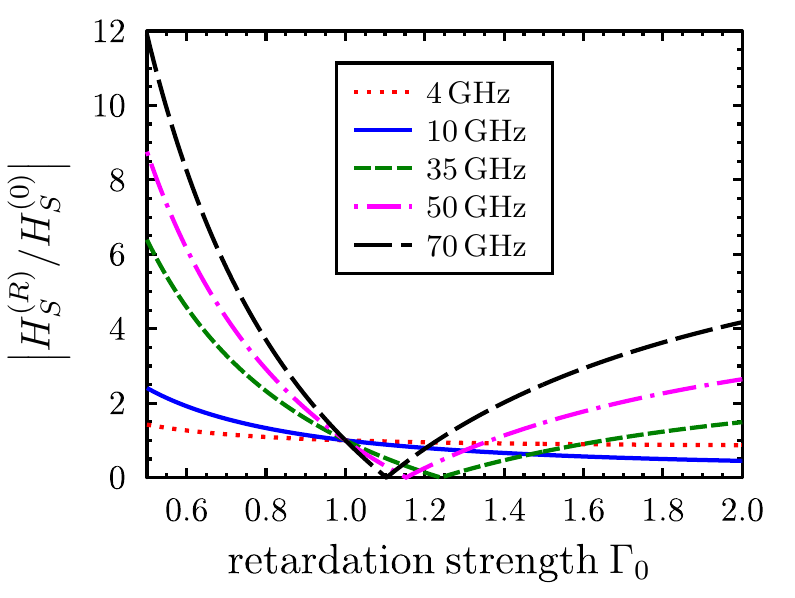}
\caption{(Color online) Effect of varying retardation strength on the uniaxial anisotropy field for various
frequencies and $\Theta_{\rm{M}}=\pi/3$.  $4\pi M_{\rm{S}}=16980\,\mathrm{G}$, $H_{\rm{S}}=-3400\,\mathrm{G},
\gamma=0.019\,\mathrm{GHz}/\mathrm{G}$, see ~\cite{Lindner:PRB80:224421:2009}.}
\label{figanisotrpy}
\ef
In this graph we assumed that all other quantities remain fixed. The effect of a varying retardation strength on the anisotropy field can clearly be seen. The change in the sign of the slope indicates that the anisotropy field $H_{\rm{S}}^{(R)}$
may even change its sign. From here we conclude that the directions of the easy axis and hard axis are interchanged. For the frequencies $4\,\mathrm{GHz}$ and $10\,\mathrm{GHz}$ this result is not observed in the range chosen for $\Gamma_0$.
Moreover, the effects become more pronounced for higher frequencies. In Fig.~\ref{figanisotrpy}
we consider only a possible alteration of the anisotropy field. Other parameters like the experimentally obtained gyromagnetic ration were unaffected. In general this parameter may also experiences a quantitative change simultaneously with $H_{\rm{S}}$.

Let us proceed by analyzing the susceptibility obtained in Eq.~\eqref{susceptibilityxx1}. Because the following discussion is referred to
the energy absorption in the film, we investigate the imaginary part of the susceptibility $\chi_{\rm{XX}}''\,$.
Since experimentally often a Lorentzian curve describes sufficiently the resonance signal we intend to arrange $\chi_{\rm{XX}}''\,$
in the form $A_0/(1+u^2)$, where $A_0$ is the absolute value of the amplitude and $u$ is a small parameter around zero. The mapping to a
Lorentzian is possible under some assumptions. Because the discussion is concentrated on the vicinity of the resonance we introduce
$\delta H=H_0-H_\mathrm{r}$, where $H_\mathrm{r}$ is the static external field when resonance occurs.
Consequently, the fields in Eq.~\eqref{H1H2} have to be replaced by $H_{1,2}\to H_{1,2}^{(r)}+\delta H\,\cos(\Theta_{\rm{H}}-\Theta_{\rm{M}})$.
Additionally, we take into account only terms of the order $\sqrt{\epsilon\lambda}$ in the final result for the linewidth where $\{\epsilon,\lambda\}\propto \{\omega/\gamma[\alpha+\omega\tau]+1/(\gamma T_2)\}$.
After a lengthy but straightforward calculation we get for $\delta H/H_{1,2}^{(\mathrm{r})}\ll1$ and using the resonance condition in
Eq.~\eqref{rescond}
\begin{align}
\begin{aligned}
\chi_{\rm{XX}}''(\omega) = \frac{A_0}{1+\left[\frac{H_0-H_\mathrm{r}}{\Delta_{\rm{T}}}\right]^2} \,,\, A_0 = \frac{M_{\rm{S}}}{(1+\kappa)\,\cos(\Theta_{\rm{H}}-\Theta_{\rm{M}})\,\Delta_{\rm{T}}} \,,\, \kappa=\frac{H_2^{(r)}}{H_1^{(r)}} \,.
\end{aligned}
\label{susceptimaginary}
\end{align}
Here we have introduced the total half-width at half-maximum (HWHM) $\Delta_{\rm{T}}$ which can be brought in the form
\begin{align}
\begin{aligned}
\Delta_{\rm{T}}=\frac{1}{\cos(\Theta_{\rm{H}}-\Theta_{\rm{M}})}\sqrt{\Delta_{\rm{G}}^2+\Delta_{\rm{B}}^2+\Delta_{\rm{GB}}^2+\Delta_{\rm{R}}^2} \,.
\end{aligned}
\label{hwhm}
\end{align}
The HWHM is a superposition of the Gilbert contribution $\Delta_{\rm{G}}$, the Bloch contribution $\Delta_{\rm{B}}$, a joint contribution $\Delta_{\rm{GB}}$ arising from
the combination of the Gilbert and Bloch damping parts in the equation of motion and the contribution $\Delta_{\rm{R}}$ which has its origin purely in the feedback mechanisms introduced into the system. The explicit expressions are
\begin{subequations}
\begin{align}
\Delta_{\rm{G}} =& \frac{\omega}{\gamma}\,\sqrt{\alpha\left[\alpha -\frac{16\sqrt{\kappa}}{(1+\kappa)}\frac{\Gamma_0\Gamma_1\,\omega\tau}{(\Gamma_0+\Gamma_1)^3}\right]} \,, \label{DeltaG} \\
\Delta_{\rm{B}} =& \frac{4\,\Gamma_0}{(\Gamma_0+\Gamma_1)}\frac{\sqrt{\kappa}}{(1+\kappa)}\,\sqrt{ \frac{1}{(\gamma T_2)^2}-\frac{4\,\Gamma_1}{(\Gamma_0+\Gamma_1)^2}\frac{\omega}{\gamma}\frac{\omega\tau}{\gamma T_2}} \,, \label{DeltaB} \\
\Delta_{\rm{GB}} =& \sqrt{\frac{ 8\Gamma_0}{(\Gamma_0+\Gamma_1)}\frac{\sqrt{\kappa}}{(1+\kappa)}\frac{\alpha\omega}{\gamma^2T_2}} \,, \label{DeltaGB} \\
\Delta_{\rm{R}} =& \frac{8\sqrt{\kappa}}{(1+\kappa)}\frac{\omega}{\gamma}\frac{\Gamma_0\Gamma_1\,\omega\tau}{(\Gamma_0+\Gamma_1)^3} \,. \label{DeltaR}
\end{align}
\end{subequations}
The parameter $\Gamma_1$ is defined in Eq.~\eqref{ftkernel}. If the expressions under the roots in Eqs.~\eqref{DeltaG} and \eqref{DeltaB}
are negative we assume that the corresponding process is deactivated and does not contribute to the linewidth $\Delta H_{\rm{T}}$. Typically,
experiments are evaluated in terms of the peak-to-peak linewidth of
the derivative $\mathrm{d}\chi_\mathrm{XX}''/\mathrm{d}H_0$, denoted as $\Delta H_\eta$. One gets
\begin{align}
\Delta H_\eta=\frac{2}{\sqrt{3}}\,\Delta_\eta \,,
\label{pplinewidth}
\end{align}
where the index $\eta$ stands for $\rm{G}$ (Gilbert contribution), $\rm{B}$ (Bloch contribution), $\rm{GB}$ (joint Gilbert-Bloch contribution), $\rm{R}$
(pure retardation contribution) or $\rm{T}$ designating the total linewidth according to Eq.~\eqref{hwhm} and Eqs.~\eqref{DeltaG}-\eqref{DeltaR}.
Obviously these equations reveal a strong nonlinear frequency dependence, which will be discussed in the subsequent section.

\section{Discussion}

As indicated in Eqs.~\eqref{hwhm} - \eqref{pplinewidth} the quantity $\Delta H_\eta$ consists of well
separated distinct contributions. The behavior of $\Delta H_\eta$ is shown in
Figs.~\ref{figDeltaHGamma0}\,-\,\ref{figDeltaHtau} as function of the three retardation parameters,
the strength $\Gamma_0$, the spatial range $\beta$ and the time scale $\tau$. In all figures the frequency
$f = \omega/(2\pi)$ is used.
\bef
\includegraphics[width=7.5cm]{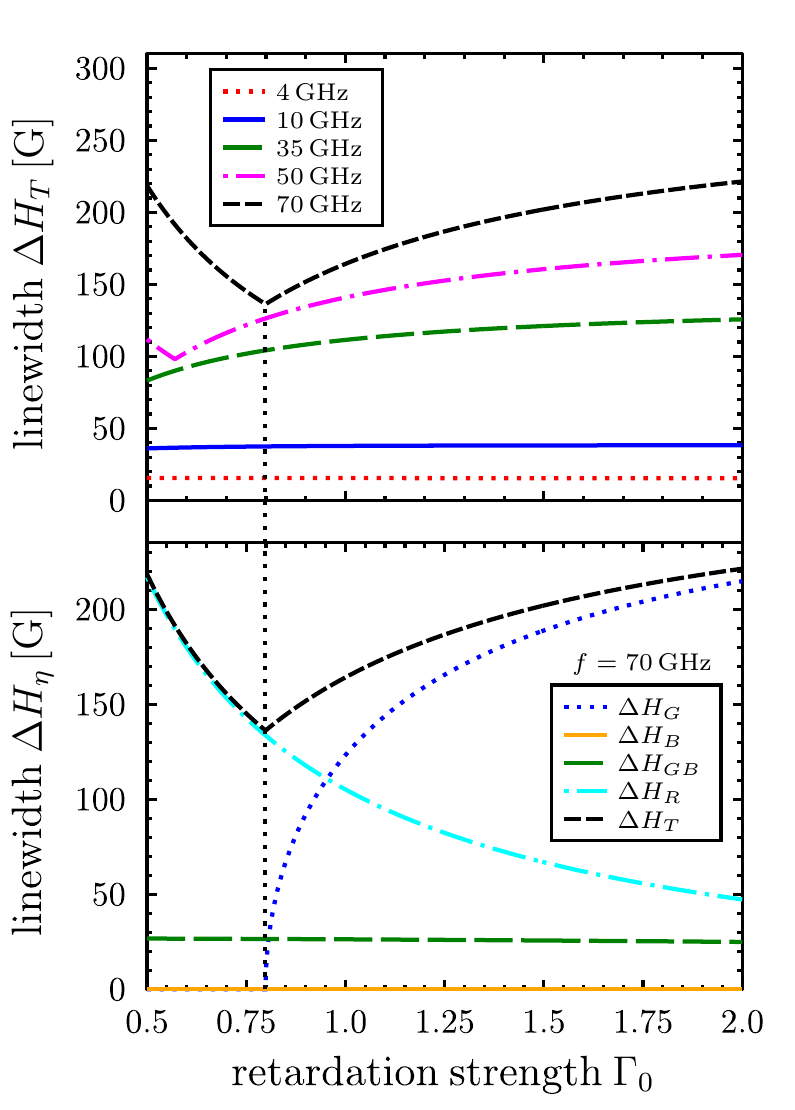}
\caption{(Color online) Influence of the retardation strength $\Gamma_0$ on the peak-to-peak linewidth $\Delta H_{\rm{T}}$ for various frequencies (top graph) and on the single contributions $\Delta H_\eta$ for $f=70\,\mathrm{GHz}$ (bottom graph). $\Delta_{\rm{B}}=0$ is this frequency region.
The parameters are: $\Theta_{\rm{H}} = \Theta_{\rm{M}}=0$, $\beta=0.5$, $\alpha=0.01$, $T_2=5\times 10^{-8}\,
\mathrm{s}, \tau=1.7\times 10^{-14}\,\mathrm{s}$. The other parameters are $4\pi M_{\rm{S}}=16980\,\mathrm{G}$, $H_{\rm{S}}=-3400\,\mathrm{G}, \gamma=0.019\,\mathrm{GHz}/\mathrm{G}$, compare ~\cite{Lindner:PRB80:224421:2009}.}
\label{figDeltaHGamma0}
\ef
In Fig.~\ref{figDeltaHGamma0} the dependence on the retardation strength $\Gamma_0$ is shown.
As already observed in Figs.~\ref{figphimphih} and \ref{figanisotrpy} a small change of $\Gamma_0$ may
lead to remarkable effects. Hence we vary this parameter in a moderate range $0.5 \leq \Gamma_0 \leq 2$.
The peak-to-peak linewidth $\Delta H_{\rm{T}}$ as function of $\Gamma_0$ remains nearly constant for
$f =4\,\mathrm{GHz}$ and $f=10\,\mathrm{GHz}$, whereas for $f=35\,\mathrm{GHz}$ a monotonous growth-up is
observed. Increasing the frequency further to $f=50\,\mathrm{GHz}$ and $70\,\mathrm{GHz}$ the curves offers
a pronounced kink. The subsequent enhancement is mainly due to the Gilbert damping. In the region of negative slope
we set $\Delta H_{\rm{G}}(\Gamma_0)=0$, while in that one with a positive slope
$\Delta H_{\rm{G}}(\Gamma_0)>0$ grows and tends to $2\,\alpha\omega /(\sqrt{3}\,\gamma)$ for $\Gamma_0\to \infty$.
The other significant contribution $\Delta H_{\rm{R}}$, arising from the retardation decay, offers
likewise a monotonous increase for growing values of the retardation parameter $\Gamma_0$. This behavior is
depicted in Fig.~\ref{figDeltaHGamma0} for $f=70\,\mathrm{GHz}$. Now let us analyze the dependence on
the dimensionless retardation length $\beta =\xi k_z$. Because $\beta$ is only nonzero if $k_z\neq0$
this parameter $\xi$ accounts the influence of excitations with nonzero wave vector. We argue that
both nonzero wave vector excitations, those arising from two-magnon scattering and those originated from
feedback mechanisms, may coincide. Based on the estimation in the previous section we consider the relevant
interval $10^{-2}\leq\beta\leq 10$. The results are shown in Fig.\ref{figDeltaHbeta}.
\bef
\includegraphics[width=7.5cm]{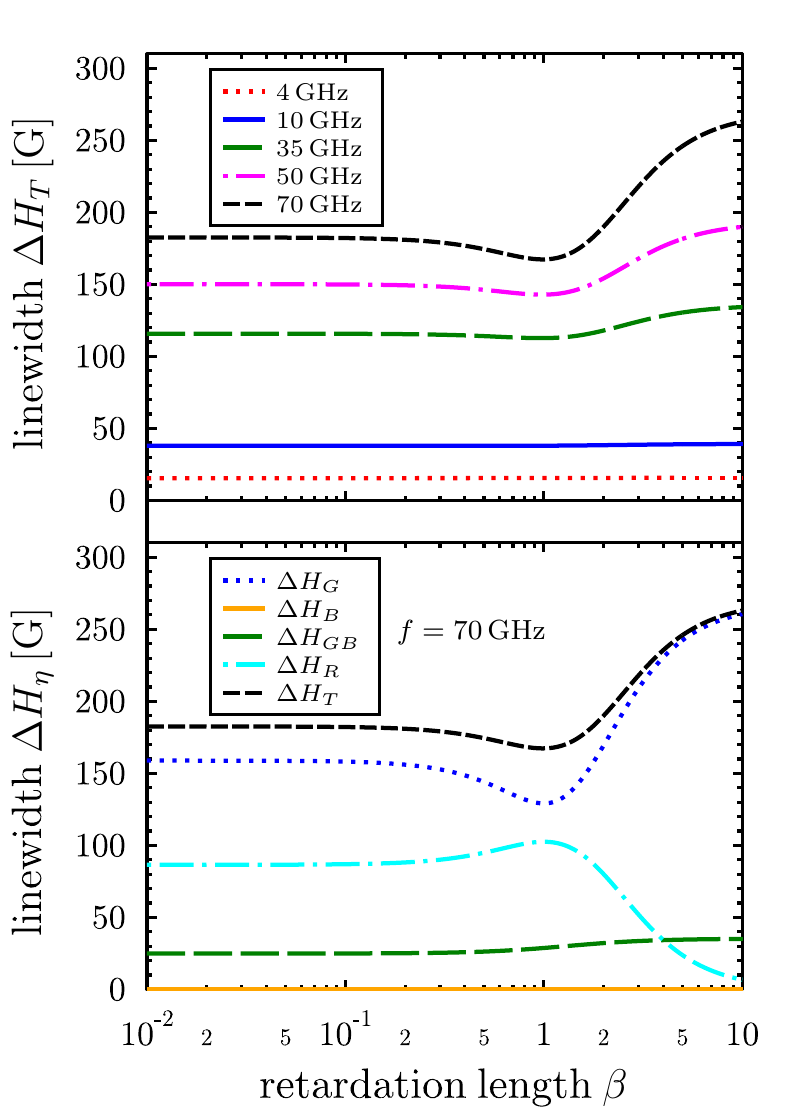}
\caption{(Color online) Influence of the dimensionless retardation length $\beta=\xi k_z$ on the total peak-to-peak linewidth $\Delta H_{\rm{T}}$ for various frequencies (top graph) and on the single contributions $\Delta H_\eta$ for $f=70\,\mathrm{GHz}$ (bottom graph); $\Delta_{\rm{B}}=0$ in this range. The parameters are:
$\Theta_{\rm{H}} = \Theta_{\rm{M}}=0$, $\Gamma_0=1.1$, $\alpha=0.01$, $T_2=5\times 10^{-8}\,\mathrm{s},
\tau=1.7\times 10^{-14}\,\mathrm{s}$. The other parameters: $4\pi M_{\rm{S}}=16980\,\mathrm{G}$, $H_{\rm{S}}=-3400\,\mathrm{G}$ and $\gamma=0.019\,\mathrm{GHz}/\mathrm{G}$ are taken from
\cite{Lindner:PRB80:224421:2009}.}
\label{figDeltaHbeta}
\ef
Within the range of $\beta$ one recognizes that the total peak-to-peak linewidths $\Delta H_{\rm{T}}$ for $f=4\,\mathrm{GHz}$ and $f=10\,\mathrm{GHz}$ offer no alteration when $\beta$ is changed. The plotted linewidths
are characterized by a minimum followed by an increase which occurs when
$\beta$ exceeds approximately $1$. This behavior is the more accentuated the larger the frequencies are.
The shape of the curve can be explained by considering the single contributions as is visible in the lower part in Fig.~\ref{figDeltaHbeta}.
While both quantities $\Delta H_{\rm{G}}(\beta)$ and $\Delta H_{\rm{R}}(\beta)$ remain constant for small $\beta$, $\Delta H_{\rm{G}}(\beta)$ tends to a minimum
and increases after that. The quantity $\Delta H_{\rm{R}}(\beta)$ develops a maximum around $\beta\approx 1$. Thus, both contributions show
nearly opposite behavior. The impact of the characteristic feedback time $\tau$ on the linewidth is illustrated in
Fig.~\ref{figDeltaHtau}.
\bef
\includegraphics[width=7.5cm]{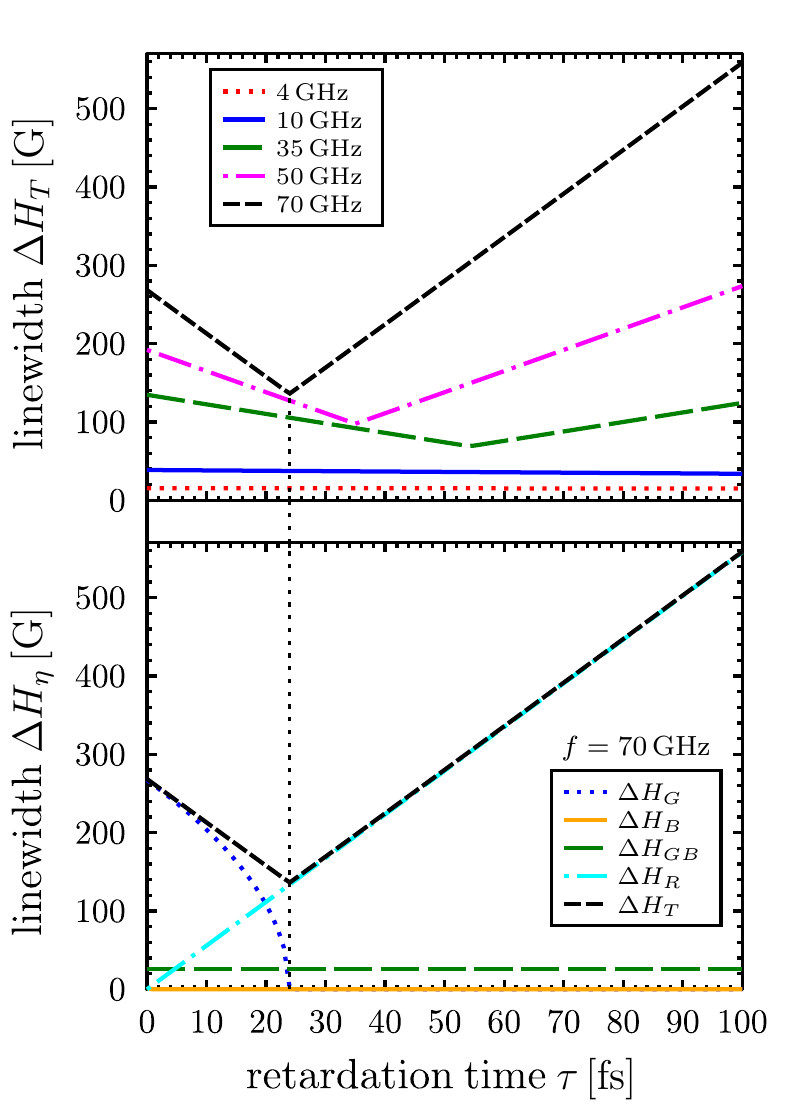}
\caption{(Color online) Influence of the retardation time $\tau$ on the total peak-to-peak linewidth $\Delta H_{\rm{T}}$ for various frequencies (top graph) and on the single contributions $\Delta H_\eta$ for $f=70\,\mathrm{GHz}$ (bottom graph). $\Delta_{\rm{B}}=0$ in this region. The parameters are
$\Theta_{\rm{H}} = \Theta_{\rm{M}}=0$, $\beta=0.5$, $\alpha=0.01$, $T_2=5\times 10^{-8}\,\mathrm{s},
\Gamma_0=1.1$; the other parameters are taken from~\cite{Lindner:PRB80:224421:2009}:~$4\pi M_{\rm{S}}=16980\,\mathrm{G}$, $H_{\rm{S}}=-3400\,\mathrm{G}, \gamma=0.019\,\mathrm{GHz}/\mathrm{G}$.}
\label{figDeltaHtau}
\ef
In this figure a linear time scale is appropriate since there are no significant effects in the
range $1\,\mathrm{fs} \geq  \tau \geq  0$.
The total linewidth $\Delta H_{\rm{T}}(\tau)$ is again nearly constant for $f=4\,\mathrm{GHz}$ and $f=10\,\mathrm{GHz}$. In contrast
$\Delta H_{\rm{T}}(\tau)$ reveals for higher frequencies two regions with differing behavior. The
total linewidth decreases until $\Delta H_{\rm{G}}(\tau)$ becomes zero. After that one observes a positive
linear slope which is due to the retardation part $\Delta H_{\rm{R}}(\tau)$.
This linear dependency is recognizable in Eq.~\eqref{DeltaR}, too. Below we will present arguments
why the feedback time $\tau$ is supposed to be in the interval $0 < \tau < 100\,\mathrm{fs}$. Before
let us study the frequency dependence of the linewidth in more detail. The general shape of the total
linewidth $\Delta H_{\rm{T}}(\omega)$ is depicted in Fig.~\ref{figlinewidthf}.
\bef
\includegraphics[width=7.5cm]{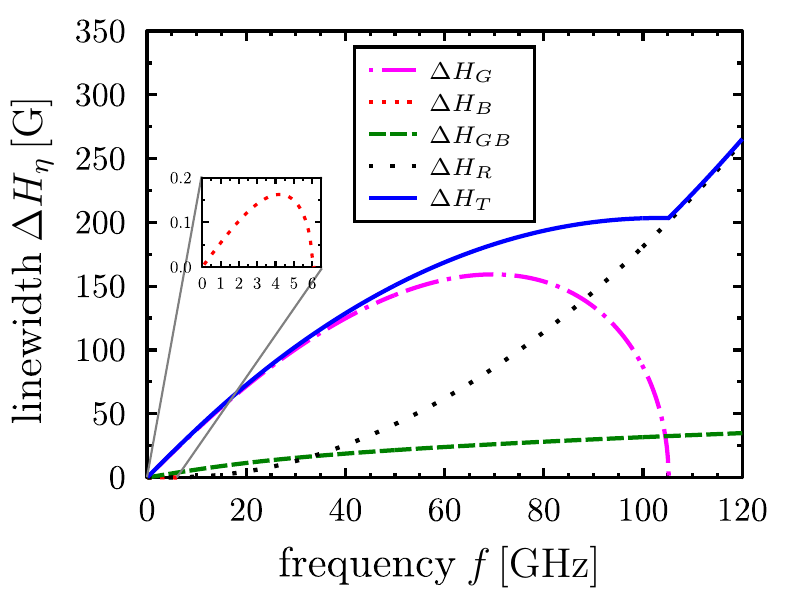}
\caption{(Color online) Frequency dependence of all contributions to the peak-to-peak linewidth for $\Theta_{\rm{H}} = \Theta_{\rm{M}}=0$, $\beta=0.5$, $\alpha=0.01$, $T_2=5\times 10^{-8}\,\mathrm{s}$, $\tau=1.7\times 10^{-14}\,\mathrm{s}$ and $\Gamma_0=1.2$.
Parameters taken from Ref.~\cite{Lindner:PRB80:224421:2009}:~$4\pi M_{\rm{S}}=16980\,\mathrm{G}$, $H_{\rm{S}}=-3400\,\mathrm{G}$ and $\gamma=0.019\,\mathrm{GHz}/\mathrm{G}$. The Bloch contribution $\Delta H_{\rm{B}}$ is shown in the inset.}
\label{figlinewidthf}
\ef
Here both the single contribution to the linewidth and the total linewidth are shown. Notice that the total
linewidth is not simply the sum of the individual contributions but has to be calculated according to
Eq.~\eqref{hwhm}. One realizes that the Bloch contribution $\Delta H_{\rm{B}}$ is only nonzero for
frequencies $f\leq 6\,\mathrm{GHz}$ in the examples shown. Accordingly $\Delta H_{\rm{B}}=0$
in Figs.~\ref{figDeltaHGamma0}-\ref{figDeltaHtau} (lower parts) since these plots refer to $f=70\,\mathrm{GHz}$.
The behavior of the Gilbert contribution deviates strongly from the typically applied linear frequency dependence.
Moreover, the Gilbert contribution will develop a maximum value and eventually it disappears at a certain
frequency where the discriminant in Eq.~\eqref{DeltaG} becomes negative. Nevertheless, the total linewidth is
a nearly monotonous increasing function of the frequency albeit, as mentioned before, for some combinations of
the model parameters there might exist a very small frequency region where $\Delta H_{\rm{G}}$ reaches zero
and the slope of $\Delta H_{\rm{T}}$ becomes slightly negative.
The loss due to the declining Gilbert part is nearly compensated or overcompensated by
the additional line broadening originated by the retardation part and the combined Gilbert-Bloch term. The latter
one is $\Delta H_{\rm{GB}} \propto \sqrt{f}$ and $\Delta H_{\rm{R}} \propto f^2$, see
Eqs.~\eqref{DeltaGB}-\eqref{DeltaR}. In the frequency region where $\Delta H_{\rm{G}}=0$ only $\Delta H_{\rm{GB}}$ and
$\Delta H_{\rm{R}}$ contribute to the total linewidth, the shape of the linewidth is mainly dominated by
$\Delta H_{\rm{R}}$. This prediction is a new result. The behavior $\Delta H_{\rm{R}}\propto f^2$, obtained in our model for high frequencies, is in contrast to conventional ferromagnetic resonance including only the sum of a
Gilbert part linear in frequency and a two-magnon contribution which is saturated at high frequencies.
So far, experimentally the frequency ranges from $1\,\mathrm{GHz}$ to $225\,\mathrm{GHz}$, see
\cite{Lenz:PhysRevB73:144424:2006}. Let us point out that the results presented in Fig.~\ref{figlinewidthf} can
be adjusted in such a manner that the Gilbert contribution will be inoperative at much higher frequencies
by the appropriate choice of the model parameters. Due to this fact we suggest an experimental verification in
more extended frequency ranges. Another aspect is the observation that excitations with a nonzero wave vector
might represent one possible retardation mechanism. Regarding Eqs.~\eqref{DeltaG}-\eqref{DeltaR} retardation
can also influence the linewidth in case $k_z=0$ (i.e. $\beta=0$ and $\Gamma_1=\Gamma_0$). Only if $\tau=0$
the retardation effects disappear. Therefore let us consider the time domain of retardation and its relation
to the Gilbert damping. The Gilbert damping and the attenuation due to retardation can be considered as competing
processes. So temporal feedback can cause that the Gilbert contribution disappears. In the same sense
the Bloch contribution is a further competing damping effect. In this regard temporal feedback has the ability
to reverse the dephasing process of spin waves based on Gilbert and Bloch damping. On the other hand the
retardation part $\Delta_{\rm{R}}$ in Eq.~\eqref{DeltaR} is always positive for $\tau>0$. Thus, the retardation
itself leads to linewidth broadening in ferromagnetic resonance and consequently to spin damping. Whether
the magnitude of retardation is able to exceed the Gilbert damping depends strongly on the frequency.
With other words, the frequency of the magnetic excitation 'decides' to which damping mechanisms the excitation
energy is transferred. Our calculation suggests that for sufficient high frequencies retardation effects dominate
the intrinsic damping behavior. Thus the orientation and the value of the magnetization within the retardation time $\tau$
plays a major role for the total damping.

Generally, experimental data should be fit according to the frequency dependence of the linewidth in terms of
Eqs.~\eqref{hwhm}-\eqref{pplinewidth}. To underline this statement we present Fig.~\ref{figvglret2mag}.
\bef
\includegraphics[width=7.5cm]{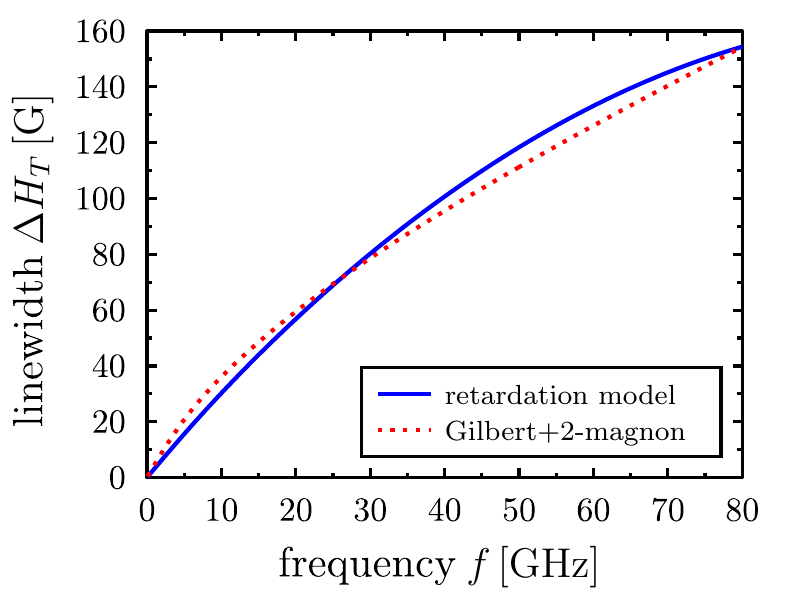}
\caption{(Color online) Comparison with the two-magnon model. Frequency dependence of the total peak-to-peak linewidth $\Delta H_{\rm{T}}$ for $\Theta_{\rm{H}} = \Theta_{\rm{M}}=0$, $\beta=0.5$, $\alpha_1=0.003$, $\alpha_2=0.0075$, $T_2=5\times 10^{-8}\,\mathrm{s}$, $\tau=1.22\times 10^{-14}\,\mathrm{s}$ and $\Gamma_0=1.2$.
Parameters taken from ~\cite{Arias:PhysRevB60:7395:1999}:~$4\pi M_{\rm{S}}=21000\,\mathrm{G}$, $H_{\rm{S}}=-15000\,\mathrm{G}$ and from ~\cite{Farle:EPL49:658:2000}: $\gamma=0.018\,\mathrm{GHz}/\mathrm{G}$ (derived from $g=2.09$ for bulk Fe). The dotted line is a superposition of Fig.~4 in
~\cite{Arias:PhysRevB60:7395:1999} reflecting the two-magnon contribution and the Gilbert contribution
(denoted as $\alpha_1$ in the text) linear in the frequency.}
\label{figvglret2mag}
\ef
In this graph we reproduce some results presented in \cite{Arias:PhysRevB60:7395:1999} for the case $\Theta_{\rm{H}}=\Theta_{\rm{M}}=0$.
To be more specific, we have used Eq.~(94) in \cite{Arias:PhysRevB60:7395:1999}
which accounts for the two-magnon scattering and the parameters given there. As result we find a copy of
Fig.~4 in \cite{Arias:PhysRevB60:7395:1999} except of the factor $2/\sqrt{3}$.
Further, we have summed up the conventional Gilbert linewidth $\propto f$ with the Gilbert damping
parameter $\alpha_1=0.003$. This superposition yields to the dotted line in Fig.~\ref{figvglret2mag}.
The result is compared with the total linewidth resulting from our retardation model plotted as solid line.
To obtain the depicted shape we set the Gilbert damping parameter according to the retardation model
$\alpha_2=0.0075$, i.e. to get a similar behavior in the same order of magnitude of $\Delta H_{\rm{T}}$ within
both approaches we have to assume that $\alpha_2$ is more than twice as large compared to $\alpha_1$.

Finally we discuss briefly the $\Theta_{\rm{H}}\,$-dependence of the linewidth which is shown in
Fig.~\ref{figDeltaHwinkel}.
\bef
\includegraphics[width=7.5cm]{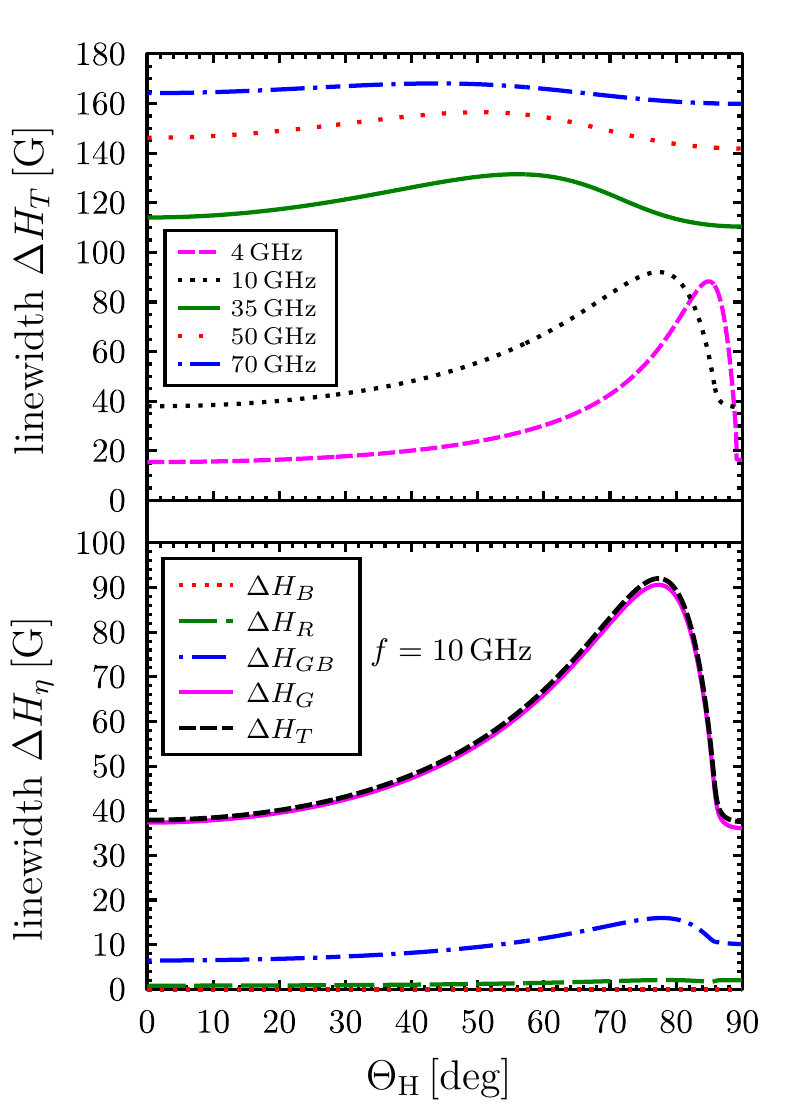}
\caption{(Color online) Angular dependence of the total peak-to-peak linewidth $\Delta H_{\rm{T}}$ for various frequencies (top graph) and all contributions
$\Delta H_\eta$ for $f=10\,\mathrm{GHz}$ (bottom graph) with $\beta=0.5$, $\alpha=0.01$, $T_2=5\times 10^{-8}\,\mathrm{s}$, $\tau=1.7\times 10^{-14}\,\mathrm{s}$ and $\Gamma_0=1.1$.
The parameters are taken from ~\cite{Lindner:PRB80:224421:2009}:~$4\pi M_{\rm{S}}=16980\,\mathrm{G}$, $H_{\rm{S}}=-3400\,\mathrm{G}$ and $\gamma=0.019\,\mathrm{GHz}/\mathrm{G}$.}
\label{figDeltaHwinkel}
\ef
In the upper part of the figure one observes that $\Delta H_{\rm{T}}(\Theta_{\rm{H}})$ exhibits a maximum which
is shifted towards lower field angles as well as less pronounced for increasing frequencies. The lower part of
Fig.~\ref{figDeltaHwinkel}, referring to $f=10\,\mathrm{GHz}$, displays that the main contribution to the
total linewidth arises from the Gilbert part $\Delta H_{\rm{G}}$. This result for $f=10\,\mathrm{GHz}$ is in
accordance with the results discussed previously, compare Fig.~\ref{figlinewidthf}. For higher frequencies
the retardation contribution $\Delta H_{\rm{R}}$ may exceed the Gilbert part.

\section{Conclusions}

A detailed study of spatiotemporal feedback effects and intrinsic damping terms offers that both mechanisms
become relevant in ferromagnetic resonance. Due to the superposition of both effects it results a nonlinear
dependence of the total linewidth on the frequency which is in accordance with experiments.
In getting the results the conventional model including Landau-Lifshitz-Gilbert damping is extended by
considering additional spatial and temporal retardation and non-conserved Bloch damping terms. Our analytical
approach enables us to derive explicit expressions for the resonance condition and the peak-to-peak linewidth.
We were able to link our results to such ones well-known from the literature. The resonance
condition is affected by the feedback strength $\Gamma_0$. The spin wave damping is likewise influenced by
$\Gamma_0$ but moreover by the characteristic memory time $\tau$ and the retardation length $\xi$. As expected
the retardation gives rise to an additional damping process. Furthermore, the complete linewidth offers a
nonlinear dependence on the frequency which is also triggered by the Gilbert damping. From here we conclude that
for sufficient high frequencies the linewidth is dominated by retardation effects. Generally, the contribution of
the different damping mechanisms to the linewidth is comprised of well separated rates which are presented in
Eqs.~\eqref{hwhm}-\eqref{pplinewidth}. Since each contribution to the linewidth is characterized by adjustable
parameters it would be very useful to verify our predictions experimentally. Notice that the contributions to
the linewidth in Eqs.~\eqref{hwhm}-\eqref{pplinewidth} depend on the shape of the retardation kernel which is
therefore reasonable not only for the theoretical approach but for the experimental verification, too.
One cannot exclude that other mechanisms as more-magnon scattering effects, nonlinear interactions,
spin-lattice coupling etc. are likewise relevant. Otherwise, we hope that our work stimulates further
experimental investigations in ferromagnetic resonance.\\

We benefit from valuable discussions about the experimental background with Dr. Khali Zakeri
from the Max-Planck-Institute of Microstructure Physics. One of us (T.B.) is grateful to the Research
Network 'Nanostructured Materials'\,, which is supported by the Saxony-Anhalt State, Germany.

\clearpage

\bibliography{FMRmemory}
\bibliographystyle{apsrev4-1}

\end{document}